\documentclass[a4paper]{article}

\usepackage{INTERSPEECH2021,amsmath,graphicx,amsfonts,multirow,booktabs,float,threeparttable,lipsum,dsfont,enumerate}

\def\x{{\mathbf{x}}}
\def\z{{\mathbf{z}}}
\def\C{{\mathbf{C}}}

\title{The DKU-DukeECE System for the Self-Supervision Speaker Verification Task of the 2021 VoxCeleb Speaker Recognition Challenge}
\name{Danwei Cai$^{1,2}$, Ming Li$^{1,2}$}

\address{
  $^1$Data Science Research Center, Duke Kunshan University, Kunshan, China\\
  $^2$Department of Electrical and Computer Engineering, Duke University, Durham, USA}
\email{\{danwei.cai, ming.li.369\}@duke.edu}

\begin{document}

\maketitle
\begin{abstract}
This report describes the submission of the DKU-DukeECE team to the self-supervision speaker verification task of the 2021 VoxCeleb Speaker Recognition Challenge (VoxSRC).
Our method employs an iterative labeling framework to learn self-supervised speaker representation based on a deep neural network (DNN).
The framework starts with training a self-supervision speaker embedding network by maximizing agreement between different segments within an utterance via a contrastive loss.
Taking advantage of DNN's ability to learn from data with label noise, we propose to cluster the speaker embedding obtained from the previous speaker network and use the subsequent class assignments as pseudo labels to train a new DNN.
Moreover, we iteratively train the speaker network with pseudo labels generated from the previous step to bootstrap the discriminative power of a DNN.
Also, visual modal data is incorporated in this self-labeling framework. The visual pseudo label and the audio pseudo label are fused with a cluster ensemble algorithm to generate a robust supervisory signal for representation learning.
Our submission achieves an equal error rate (EER) of 5.58\% and 5.59\% on the challenge development and test set, respectively.
\end{abstract}

\section{Introduction}

This report describes the submission of the DKU-DukeECE team to the self-supervision speaker verification task of the 2021 VoxCeleb Speaker Recognition Challenge (VoxSRC).

In our previous work on self-supervised speaker representation learning \cite{cai_iterative_2021}, we proposed a two-stage iterative labeling framework. In the first stage, contrastive self-supervised learning is used to pre-training the speaker embedding network. This allows the network to learn a meaningful feature representation for the first round of clustering instead of random initialization. In the second stage, a clustering algorithm iteratively generates pseudo labels of the training data with the learned representation, and the network is trained with these labels in a supervised manner. The clustering algorithm can discover the intrinsic structure of the representation of the unlabeled data, providing meaningful supervisory signals comparing to contrastive learning which draws negative samples uniformly from the training data without label information. The idea behind the proposed framework is to take advantage of the DNN's ability to learn from data with label noise and bootstrap its discriminative power.

In this work, we extend this iterative labeling framework to multi-modal audio-visual data, considering that complementary information from different modalities can help the clustering algorithm generate more meaningful supervisory signals. Specifically, we train a visual representation network to encode face information using the pseudo labels generated by audio data. With the resulted visual representations, clustering is performed to generate pseudo labels for visual data. Then, we employ a cluster ensemble algorithm to fuse pseudo-labels generated by different modalities. This fused pseudo-label is then used to train speaker and face representation networks. With the clustering ensemble algorithm, information in one modality can flow to the other modality, providing more robust and fault-tolerant supervisory signals.

\begin{figure*}[t]
\centering
  \label{fig:fc2}
  \includegraphics[width=\linewidth]{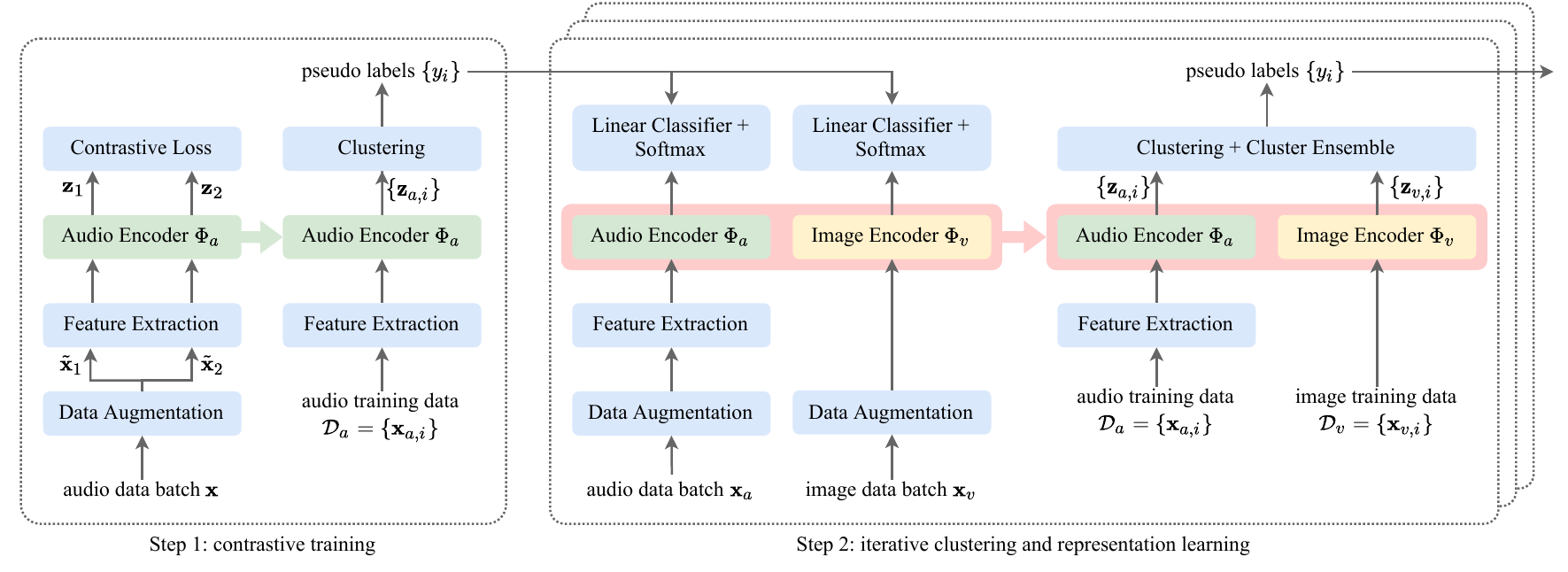}
  \caption{The proposed iterative framework for self-supervised speaker representation learning using multi-modal data.}
\end{figure*}

\section{Methods}
This section describes the proposed iterative labeling framework for self-supervised speaker representation learning using multi-modal audio-visual data. We illustrate the proposed framework in figure 1.
\begin{itemize}
    \item Stage 1: contrastive training
    \begin{itemize}
    \item Train an audio encoding network using contrastive self-supervised learning.
    \item With this encoding network, extract representations for the audio data. Perform a clustering algorithm on these audio representations to generate pseudo labels.
    \end{itemize}
    \item Stage 2: iterative clustering and representation learning
    \begin{itemize}
    \item With the generated pseudo labels, train audio and visual encoding network independently in a supervised manner.
    \item With the audio encoding network, extract audio representations and perform clustering to generate audio pseudo labels.
    \item With the visual encoding network, extract visual representations and perform clustering to generate visual pseudo labels.
    \item Fuse the audio and visual pseudo labels using a cluster ensemble algorithm.
    \item Repeat stage 2 with limited rounds.
    \end{itemize}
\end{itemize}

\subsection{Contrastive self-supervised learning}
We employ the contrastive self-supervised learning (CSL) framework similar to the framework in \cite{chen_simple_2020, falcon_framework_2020} to learn an initial audio representatoion.
Let $\mathcal{D} = \{\x_1, \cdots, \x_N\}$ be an unlabeled dataset with $N$ data samples, CSL assumes that each data sample defines its own class and perform instance discrimination.
During training, we randomly sample a mini-batch $\mathcal{B} = \{\x_1, \cdots, \x_M\}$ of $M$ data samples from $\mathcal{D}$.
For data point $\x_i$, stochastic data augmentation is performed to generate two correlated views, i.e., $\tilde{\x}_{i,1}$ and $\tilde{\x}_{i,2}$, resulting $2M$ data points in total for a mini-batch. Two different audio segments are randomly cropped from the original audio before data augmentation. $\tilde{\x}_{i,1}$ and $\tilde{\x}_{i,2}$ are considered as a positive pair and other $2(M-1)$ data points $\{\tilde{\x}_{j,k} | j\neq i, k=1,2\}$ are negative examples for $\tilde{\x}_{i,1}$ and $\tilde{\x}_{i,2}$.

During training, a neural network encoder $\Phi$ extracts representations for the $2M$ augmented data samples,
\begin{equation}
    \z_{i,j} = \Phi (\tilde{\x}_{i,k}), \ k\in\{1,2\}
\end{equation}

After that, contrastive loss identifies the positive example $\tilde{\x}_{i,1}$ (or $\tilde{\x}_{i,2}$) among the negative examples $\{\tilde{\x}_{j,k} | j\neq i, k=1,2\}$ for $\tilde{\x}_{i,2}$ (or $\tilde{\x}_{i,1}$). We adapt the contrastive loss from SimCLR~\cite{chen_simple_2020} as:
\begin{equation}
\mathcal{L}_\mathrm{CSL} = \frac{1}{2M} \sum_{i=1}^M (l_{i,1} + l_{i,2})
\end{equation}
\begin{equation}
l_{i,j} = - \log \frac{\exp(\cos(\z_{i,1}, \z_{i,2})/\tau)}{\sum_{k=1}^M\sum_{l=1}^2 \mathds{1}_{\substack{k \neq i\\l \neq j}}\exp(\cos(\z_{i,j}, \z_{k,l})/\tau)}
\label{equation:csl}
\end{equation}
where $\mathds{1}$ is an indicator function evaluating $1$ when $k \neq i$ and  $l \neq j$, $\cos$ denotes the cosine similarity and $\tau$ is a temperature parameter to scale the similarity scores.
$l_{i,j}$ can be interpreted as the loss for anchor feature $\z_{i,j}$.
It computes positive score for positive feature $\z_{i,(j+1)\mathrm{mod}2}$ and negative scores across all $2(M-1)$ negative features $\{\z_{k,j} | k\neq i, j=1,2\}$.

\begin{figure}[t]
\centering
  \includegraphics[width=\linewidth]{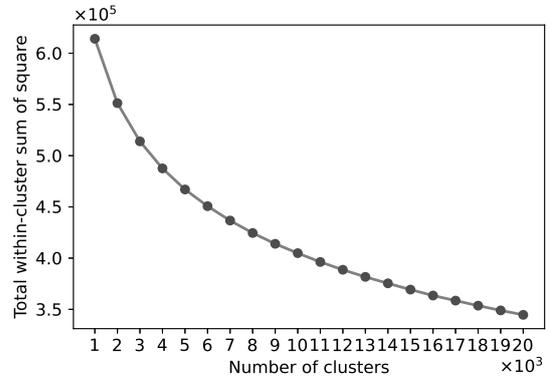}
  \caption{Within-cluster sum of square $W$ of a clustering procedure versus the number of clusters $K$ employed.}
  \label{fig:wss}
\end{figure}

\subsection{Generating pseudo labels by clustering}

\subsubsection{K-means clustering}

Given the learned representations of the training data, we employ a clustering algorithm to generate cluster assignments and pseudo labels. In this paper, we use the well-known \textit{k}-means algorithm because of its simplicity, fast speed, and capability with large datasets.

Let the learnt representation in $d$-dimensional feature space $\z\in\mathbb{R}^d$, \textit{k}-means learns a centroid matrix $\C\in \mathbb{R}^{d\times K}$ and the cluster assignment $y_i \in \{1,\cdots,K\}$ for representation $\z_i$ with the following learning objective
\begin{equation}
    \min_{\C}\frac{1}{N}\sum_{i=1}^{N}\min_{y_i} \| \z_i -\C_{y_i}\|^2_2
\end{equation}
where $\C_{y_i}$ is the $y_i^\text{th}$ column of the centroid matrix $\C$. The optimal assignments $\{y_1, \cdots, y_N\}$ are used as pseudo labels.

\subsubsection{Determine the number of clusters}

To determine the optimal number of clusters, we employ the simple `elbow' method. It calculates the total within-cluster sum of squares $W$ for the clustering outputs with different numbers of clusters $K$:
\begin{equation}
    W = \sum_{i=1}^{N} \| \z_i -\C_{y_i}\|^2_2
\end{equation}
The curve of the total within-cluster sum of squares $W$ is plotted according to a sequence of $K$ in ascending order. Figure \ref{fig:wss} shows an example of such a curve. $W$ decreases as $K$ increases, and the decrease of $W$ flattens from some $K$ onwards, forming an `elbow' of the curve. Such `elbow' indicates that additional clusters beyond such $K$ contribute little intra-cluster variation; thus, the $K$ at the `elbow' indicates the appropriate number of clusters. In figure \ref{fig:wss}, the number of clusters can choose between 5000 to 7000.

This `elbow' method is not exact, and the choice of the optimal number of clusters can be subjective. Still, it provides a meaningful way to help to determine the optimal number of clusters. A mathematically rigorous interpretation of this method can be found in \cite{tibshirani_estimating_2001}.

\subsection{Learning with pseudo labels}
Given a multi-modal dataset with audio-modality $\mathcal{D}_a = \{\x_{a,1}, \\ \x_{a,2}, \cdots, \x_{a,N}\}$ and visual-modality $\mathcal{D}_v=\{\x_{v,1}, \x_{v,2}, \cdots, \\ \x_{v,N}\}$, an audio encoder $\Phi_a$ and a visual encoder $\Phi_v$ are discriminatively trained with an audio classifier $g_{W_a}(\cdot)$ and a visual classifier $g_{W_v}(\cdot)$ respectively using the generated pseudo labels $\{y_1, \cdots, y_{N}\}$. For each modality, the representation can be extracted as 
\begin{equation}\begin{aligned}
    \z_a &= \Phi_a (\x_a) \\
    \z_v &= \Phi_v (\x_v)
\end{aligned}\end{equation}

For a single modality, the parameters $\{\Phi_a, W_a\}$ or $\{\Phi_v, W_v\}$ are jointly trained with the cross-entropy loss:
\begin{equation}
\label{loss_cross_entropy}
    \mathcal{L}_\mathrm{classifier} = - \sum_{i=1}^{N} \sum_{k=1}^{K}\log\left(p(k|\x_i)q(k|\x_i)\right)
\end{equation}
\begin{equation}
    p(k|\x_i) = \frac{\exp({g_{W}}_k(\z_i))}{\sum_{j=1}^{K}\exp({g_{W}}_{j}(\z_i))}
\end{equation}
where $q(k|\x_i) = \delta_{k, y_i}$ is the ground-truth distribution over labels for data sample $\x_i$ with label $y_i$, $\delta_{k, y_i}$ a Dirac delta which equals to $1$ for $k=y_i$ and 0 otherwise, ${g_{W}}_{j}(\z_i)$ is the $j^\mathrm{th}$ element ($j\in\{1,\cdots,K\}$) of the class score vector ${g_{W}}(\z_i)\in \mathbb{R}^K$, $K$ is the number of the pseudo classes.

\subsection{Clustering audio-visual data}
Clustering on the audio representations $\{\z_{a,i} | i=1,\cdots, N\}$ and the visual representations $\{\z_{v,i} | i=1,\cdots, N\}$ gives audio pseudo labels $\{y_{a,i} | i=1,\cdots, N\}$ and visual pseudo labels $\{y_{v,i} | i=1,\cdots, N\}$ respectively.

Considering that the audio and the visual representations contain complementary information from different modalities, we apply an additional clustering on the joint representations to generate more robust pseudo labels. Given the audio representation $\z_a$ and the visual representation $\z_v$, concatenating $\z_a$ and $\z_v$ gives the joint representation $\z_\jmath=(\z_a,\z_v)$. The pseudo labels $\{y_{\jmath,i} | i=1,\cdots, N\}$ is then generated by clustering on joint representations.

\subsection{Cluster ensemble}
We use simple voting strategy \cite{bauer1999empirical,lam1997application} to fuse the three clustering outputs, i,e., $\{y_{a,i}\}$, $\{y_{v,i}\}$ and $\{y_{\jmath,i}\}$. Since the cluster labels in different clustering outputs are arbitrary, cluster correspondence should be established among different clustering outputs. This starts with a contingency matrix $\Omega\in\mathbb{R}^{K\times K}$ for the referenced clustering output $\{y_{\text{ref},i}\}$ and the current clustering output $\{y_{\text{cur},i}\}$, where $K$ is the number of clusters. Each entry $\Omega_{l, l'}$ represents the co-occurence between cluster $l$ of the referenced clustering output and cluster $l'$ of the current clustering output,
\begin{equation}
	\Omega_{l,l'}=\sum_{i=1}^{N} \omega(i), \ \ 
	\omega(i)=\begin{cases} 1 & y_{\text{ref},i} = l,  y_{\text{cur},i} = l' \\ 0 & \text{otherwise} \end{cases}
\end{equation}
Cluster correspondence is solved by the following optimization problem,
\begin{equation}
	\max_\Theta \sum_{l=1}^{K} \sum_{l'=1}^{K} \Omega_{l,l'}\Theta_{l,l'}
\end{equation}
where $\Theta\in\mathbb{R}^{K\times K}$ is the correspondence matrix for the two clustering outputs. $\Theta_{l,l'}$ equals to $1$ if cluster $l$ in the reference clustering output corresponds to cluster $l'$ in the current clustering output, $0$ otherwise. This optimization can be solved by the Hungarian algorithm \cite{munkres1957algorithms}.

We select the joint pseudo labels as the reference clustering output and calculate cluster correspondence for the audio and visual pseudo labels. A globally consistent label set is obtained after the re-labeling process. Majority voting is then employed to determine a consensus pseudo label for each data sample in the multi-modal dataset.

\subsection{Dealing with label noise: label smoothing regularization}
One problem with the generated pseudo labels is label noise which degrades the generalization performance of deep neural networks. We apply label smoothing regularization to estimate the marginalized effect of label noise during training. It prevents a DNN from assigning full probability to the training samples with noisy label \cite{szegedy_rethinking_2016, pereyra_regularizing_2017}. Specifically, for a training example $\x$  with label $y$, label smoothing regularization replaces the label distribution $q(k|\x)=\delta_{k, y}$ in equation (\ref{loss_cross_entropy}) with
\begin{equation}
	q'(k|\x) = (1-\epsilon)\delta_{k,y} + \frac{\epsilon}{K}
\end{equation}
where $\epsilon$ is a smoothing parameter and is set to $0.1$ in the experiments.

\section{Experiments}

\subsection{Dataset}
The experiments are conducted on the development set of Voxceleb 2, which contains 1,092,009 video segments from 5,994 individuals \cite{chung_voxceleb2:_2018}. Speaker labels are not used in the proposed method. For evaluation, the development set and test set of Voxceleb 1 are used \cite{nagrani_voxceleb:_2017}. For each video segment in VoxCeleb datasets, we extracted image six frames per second.

\subsection{Data augmentation}
\subsubsection{Data augmentation for audio data}
Data augmentation is proven to be an effective strategy for both conventional learning with supervision \cite{cai_within-sample_2020} and contrastive self-supervision learning \cite{inoue_semi-supervised_2020,huh_augmentation_2020,chen_simple_2020} in the context of deep learning. We perform data augmentation with MUSAN dataset \cite{musan}. The noise type includes ambient noise, music, and babble noise for the background additive noise. The babble noise is constructed by mixing three to eight speech files into one. For the reverberation, the convolution operation is performed with 40,000 simulated room impulse responses (RIR) in MUSAN. We only use RIRs from small and medium rooms.

With contrastive self-supervised learning, three augmentation types are randomly applied to each training utterance: applying only noise addition, applying only reverberation, and applying both noise and reverberation. The signal-to-noise ratios (SNR) are set between 5 to 20 dB.

When training with pseudo labels, data augmentation is performed at a probability of 0.6. The SNR is randomly set between 0 to 20 dB.

\begin{table*}[t]
  \caption{Speaker verification performance (EER[\%]) on Voxceleb 1 test set. The NMIs of the pseudo labels for each iteration are also reported.}
  \label{tab:1}
  \centering
  \begin{tabular}[c]{l|cc|cc|c}
    \toprule
    Model & audio NMI & audio EER & visual NMI & visual EER & fused label NMI \\
    \midrule
    Fully supervised & 1 & 1.51 & -& -& - \\
    \midrule
    Initial round (CSL) & 0.75858 & 8.86 & - & - & - \\
    Round 1 & 0.90065 & 3.64 & 0.91071 & 5.55 & 0.95053 \\
    Round 2 & 0.94455 & 2.05 & 0.95017 & 2.27 & 0.95739 \\
    Round 3 & 0.95196 & 1.93 & 0.95462 & 1.78 & 0.95862 \\
    Round 4 & - & 1.81 & - & - & - \\
    \bottomrule
  \end{tabular}
\end{table*}

\begin{table}[t]
  \caption{Speaker verification performance on VoxSRC 2021 development and test set.}
  \label{tab:2}
  \centering
  \begin{tabular}[c]{l|cc|cc}
    \toprule
     & \multicolumn{2}{c}{original score} & \multicolumn{2}{c}{after score norm} \\
     & minDCF & EER[\%] & minDCF & EER[\%]\\
    \midrule
    System 1 & 0.386 & 6.310 & 0.341 & 6.214 \\
	System 2 & 0.375 & 6.217 & 0.336 & 6.057 \\
	System 3 & 0.392 & 6.224 & 0.361 & 6.067 \\
	Fusion   & 0.344 & 5.683 & 0.315 & 5.578 \\
	\midrule
	Fusion (test) & - & - & 0.341 & 5.594 \\
    \bottomrule
  \end{tabular}
\end{table}

\subsubsection{Data augmentation for visual data}
We sequentially apply these simple augmentations for the visual data: random cropping followed by resizing to $224\times224$, random horizontal flipping, random color distortions, random grey scaling, and random Gaussian blur. The data augmentation is performed at a probability of 0.6. We normalize each image's pixel value to the range of $[-0.5, 0.5]$ afterward.

\subsection{Network architecture}
\subsubsection{Audio encoder}
We opt for a residual convolutional network (ResNet) to learn speaker representation from the spectral feature sequence of varying length \cite{cai_exploring_2018}. The ResNet's output feature maps are aggregated with a global statistics pooling layer, which calculates means and standard deviations for each feature map. A fully connected layer is employed afterward to extract the 128-dimensional speaker embedding.
\subsubsection{Visual encoder}
We choose the standard ResNet-34 \cite{He2016Deep} as the visual encoder. After the pooling layer, a fully connected layer transforms the output to a 128-dimensional embedding.

\subsection{Implementation details}

\subsubsection{Contrastive self-supervised learning setup}
We choose a 40-dimensional log Mel-spectrogram with a 25ms Hamming window and 10ms shifts for audio data for feature extraction. The duration between 2 to 4 seconds is randomly generated for each audio data batch.

We use the same network architecture as in \cite{cai_within-sample_2020} but with half feature map channels. ReLU non-linear activation and batch normalization are applied to each convolutional layer in ResNet. Network parameters are updated using Adam optimizer \cite{kingma_adam_2017} with an initial learning rate of 0.001 and a batch size of 256. The temperature $\tau$ in equation (\ref{equation:csl}) is set as 0.1.

\subsubsection{Clustering setup}
The cluster number is set to 6,000 for \textit{k}-means based on the `elbow' method described in section 2.2.2. The $W$-$K$ curve shown in figure \ref{fig:wss} is based on the audio representation learned with contrastive loss. With the dataset size of 100,000, we range the number of clusters $K$ from 1,000 to 20,000, considering the average cluster size ranging from 50 to 1,000.

\subsubsection{Setup for supervised training}
For the audio data, we extract 80-dimensional log Mel-spectrogram as input features. The duration between 2 to 4 seconds is randomly generated for each audio data batch. The architecture of the audio encoder is the same as the one used in\cite{cai_within-sample_2020}.

For both audio and visual encoders, dropout is added before the classification layer to prevent overfitting \cite{srivastava_dropout:_2014}. Network parameters are updated using the stochastic gradient descent (SGD) algorithm. The learning rate is initially set to 0.1 and is divided by ten whenever the training loss reaches a plateau.

\subsection{Robust training on final pseudo labels}
Our final submission consists of three systems trained on the final pseudo labels.
\begin{itemize}
	\item System 1: the network architecture is the same as the self-labeling framework; label smoothing regularization is applied.
	\item System 2: Squeeze-Excitation (SE) module \cite{hu_squeeze-and-excitation_2019} is added to the network in the self-labeling framework, AAM-softmax \cite{deng_arcface_2019} loss is used to train the network.
	\item System 3: same as system 2; the single-speaker audio segment information is used to improve the final pseudo label: the mode label is used as the final label of the single-speaker audio segment.
\end{itemize}

\subsection{Score normalization}
After scoring with cosine similarity, scores from all trials are subject to score normalization. We utilize Adaptive Symmetric Score Normalization (AS-Norm) in our systems \cite{matejka_analysis_2017}. The number of the cohort is 300 for all systems.

\subsection{Experimental results}
Table \ref{tab:1} shows the results of each clustering iteration on Voxceleb 1 original test set. Normalized mutual information (NMI) is used as a measurement of clustering quality. With four rounds of training, our method obtains an EER of 1.81\%.

Table \ref{tab:2} shows the results of our submission system on the VoxSRC 2021 development and test set.

\bibliographystyle{IEEEtran}

\bibliography{mybib}

\begin{thebibliography}{10}
\providecommand{\url}[1]{#1}
\csname url@samestyle\endcsname
\providecommand{\newblock}{\relax}
\providecommand{\bibinfo}[2]{#2}
\providecommand{\BIBentrySTDinterwordspacing}{\spaceskip=0pt\relax}
\providecommand{\BIBentryALTinterwordstretchfactor}{4}
\providecommand{\BIBentryALTinterwordspacing}{\spaceskip=\fontdimen2\font plus
\BIBentryALTinterwordstretchfactor\fontdimen3\font minus
  \fontdimen4\font\relax}
\providecommand{\BIBforeignlanguage}[2]{{%
\expandafter\ifx\csname l@#1\endcsname\relax
\typeout{** WARNING: IEEEtran.bst: No hyphenation pattern has been}%
\typeout{** loaded for the language `#1'. Using the pattern for}%
\typeout{** the default language instead.}%
\else
\language=\csname l@#1\endcsname
\fi
#2}}
\providecommand{\BIBdecl}{\relax}
\BIBdecl

\bibitem{cai_iterative_2021}
D.~Cai, W.~Wang, and M.~Li, ``An {{Iterative Framework}} for
  {{Self}}-{{Supervised Deep Speaker Representation Learning}},'' in
  \emph{ICASSP}, 2021, pp. 6728--6732.

\bibitem{chen_simple_2020}
T.~Chen, S.~Kornblith, M.~Norouzi, and G.~Hinton, ``A {{Simple Framework}} for
  {{Contrastive Learning}} of {{Visual Representations}},'' in \emph{{{ICML}}},
  2020, pp. 1597--1607.

\bibitem{falcon_framework_2020}
W.~Falcon and K.~Cho, ``A {{Framework For Contrastive Self}}-{{Supervised
  Learning And Designing A New Approach}},'' \emph{arXiv:2009.00104}, 2020.

\bibitem{tibshirani_estimating_2001}
R.~Tibshirani, G.~Walther, and T.~Hastie, ``Estimating the number of clusters
  in a data set via the gap statistic,'' \emph{Journal of the Royal Statistical
  Society: Series B (Statistical Methodology)}, vol.~63, no.~2, pp. 411--423,
  2001.

\bibitem{bauer1999empirical}
E.~Bauer and R.~Kohavi, ``{An Empirical Comparison of Voting Classification
  Algorithms: Bagging, Boosting, and Variants},'' \emph{Machine learning},
  vol.~36, no.~1, pp. 105--139, 1999.

\bibitem{lam1997application}
L.~Lam and S.~Suen, ``{Application of Majority Voting to Pattern Recognition:
  An Analysis of Its Behavior and Performance},'' \emph{IEEE Transactions on
  Systems, Man, and Cybernetics-Part A: Systems and Humans}, vol.~27, no.~5,
  pp. 553--568, 1997.

\bibitem{munkres1957algorithms}
J.~Munkres, ``{Algorithms for the Assignment and Transportation Problems},''
  \emph{Journal of the society for industrial and applied mathematics}, vol.~5,
  no.~1, pp. 32--38, 1957.

\bibitem{szegedy_rethinking_2016}
C.~Szegedy, V.~Vanhoucke, S.~Ioffe, J.~Shlens, and Z.~Wojna, ``Rethinking the
  {{Inception Architecture}} for {{Computer Vision}},'' in \emph{{{CVPR}}},
  2016, pp. 2818--2826.

\bibitem{pereyra_regularizing_2017}
G.~Pereyra, G.~Tucker, J.~Chorowski, {\L}.~Kaiser, and G.~Hinton,
  ``Regularizing {{Neural Networks}} by {{Penalizing Confident Output
  Distributions}},'' in \emph{{{ICLR}} ({{Workshop}})}, 2017.

\bibitem{chung_voxceleb2:_2018}
J.~S. Chung, A.~Nagrani, and A.~Zisserman, ``Voxceleb2: {{Deep Speaker
  Recognition}},'' in \emph{Interspeech}, 2018, pp. 1086--1090.

\bibitem{nagrani_voxceleb:_2017}
A.~Nagrani, J.~S. Chung, and A.~Zisserman, ``Voxceleb: {A} {Large}-{Scale}
  {Speaker} {Identification} {Dataset},'' in \emph{Interspeech}, 2017, pp.
  2616--2620.

\bibitem{cai_within-sample_2020}
D.~Cai, W.~Cai, and M.~Li, ``Within-{Sample} {Variability}-{Invariant} {Loss}
  for {Robust} {Speaker} {Recognition} {Under} {Noisy} {Environments},'' in
  \emph{{ICASSP}}, 2020, pp. 6469--6473.

\bibitem{inoue_semi-supervised_2020}
N.~Inoue and K.~Goto, ``Semi-{{Supervised Contrastive Learning}} with
  {{Generalized Contrastive Loss}} and {{Its Application}} to {{Speaker
  Recognition}},'' \emph{arXiv:2006.04326}, 2020.

\bibitem{huh_augmentation_2020}
J.~Huh, H.~S. Heo, J.~Kang, S.~Watanabe, and J.~S. Chung, ``{Augmentation
  Adversarial Training for Unsupervised Speaker Recognition},''
  \emph{arXiv:2007.12085}, 2020.

\bibitem{musan}
D.~Snyder, G.~Chen, and D.~Povey, ``{MUSAN}: {A} {Music}, {Speech}, and {Noise}
  {Corpus},'' \emph{arXiv:1510.08484}, 2015.

\bibitem{cai_exploring_2018}
W.~Cai, J.~Chen, and M.~Li, ``Exploring the {Encoding} {Layer} and {Loss}
  {Function} in {End}-to-{End} {Speaker} and {Language} {Recognition}
  {System},'' in \emph{Speaker Odyssey}, 2018, pp. 74--81.

\bibitem{He2016Deep}
K.~He, X.~Zhang, S.~Ren, and J.~Sun, ``{Deep Residual Learning for Image
  Recognition},'' in \emph{CVPR}, 2016, pp. 770--778.

\bibitem{kingma_adam_2017}
D.~P. Kingma and J.~Ba, ``Adam: {{A Method}} for {{Stochastic Optimization}},''
  in \emph{{ICLR}}, 2015.

\bibitem{srivastava_dropout:_2014}
N.~Srivastava, G.~Hinton, A.~Krizhevsky, I.~Sutskever, and R.~Salakhutdinov,
  ``Dropout: {{A Simple Way}} to {{Prevent Neural Networks}} from
  {{Overfitting}},'' \emph{Journal of Machine Learning Research}, vol.~15,
  no.~1, pp. 1929--1958, 2014.

\bibitem{hu_squeeze-and-excitation_2019}
J.~Hu, L.~Shen, S.~Albanie, G.~Sun, and E.~Wu, ``Squeeze-and-{{Excitation
  Networks}},'' \emph{CVPR}, 2019.

\bibitem{deng_arcface_2019}
J.~Deng, J.~Guo, N.~Xue, and S.~Zafeiriou, ``{{ArcFace}}: {{Additive Angular
  Margin Loss}} for {{Deep Face Recognition}},'' in \emph{{{CVPR}}}, 2019, pp.
  4685--4694.

\bibitem{matejka_analysis_2017}
P.~Mat{\v e}jka, O.~Novotn{\'y}, O.~Plchot, L.~Burget, M.~D. S{\'a}nchez, and
  J.~{\v C}ernock{\'y}, ``Analysis of {{Score Normalization}} in {{Multilingual
  Speaker Recognition}},'' in \emph{The {{Annual Conference}} of the
  {{International Speech Communication Association}} ({{INTERSPEECH}})}, 2017,
  pp. 1567--1571.

\end{thebibliography}

\end{document}